\documentclass{article}
\usepackage[utf8]{inputenc}
\usepackage[english]{babel}
\usepackage{authblk}
\usepackage{amsmath}
\usepackage{csquotes}
\usepackage{float}
\usepackage{rotating}
\usepackage[a4paper, total={6in, 8in}]{geometry}
\usepackage{epsfig}
\usepackage[sorting=none]{biblatex}
\usepackage{chemformula}
\usepackage{hyperref}

\usepackage{setspace}
\title{A First-Principles Study on Electronic, Thermodynamic, and Dielectric Properties of Monolayer \ch{Ca(OH)_{2}} and \ch{Mg(OH)_{2}}}

\author[1]{Mehrdad Rostami Osanloo}
\author[2]{Kolade A. Oyekan}
\author[2]{William G. Vandenberghe$^{*}$}
\affil[1]{Department of Physics, University of Texas at Dallas, Richardson, TX, USA}
\affil[2]{Department of Materials Science and Engineering, University of Texas at Dallas, Richardson, TX, USA}

\date{March 2022}

\begin{document}

\maketitle

\begin{abstract}

\noindent We perform first-principles calculations to explore electronic, thermodynamic, and dielectric properties of two-dimensional (2D) layered, alkaline-earth hydroxides \ch{Ca(OH)_{2}} and \ch{Mg(OH)_{2}}. We calculate the lattice parameters, exfoliation energies, and phonon spectra of monolayers and also investigate the thermal properties of these monolayers such as Helmholtz free energy, heat capacity at constant volume, and entropy as a function of temperature. We employ Density Functional Perturbation Theory (DFPT) to calculate the in-plane and out-of-plane static dielectric constant of the bulk and monolayer samples. We compute the bandgap and electron affinity values using the HSE06 functional and estimate the leakage current density of transistors with monolayer \ch{Ca(OH)_{2}} and \ch{Mg(OH)_{2}} as dielectrics when combined with \ch{HfS_{2}} and \ch{WS_{2}}, respectively. Our results show that bilayer \ch{Mg(OH)_{2}} (EOT $\sim$ 0.60 nm) with a lower solubility in water, offers higher out-of-plane dielectric constants and lower leakage currents than bilayer \ch{Ca(OH)_{2}} (EOT $\sim$ 0.56 nm). Additionally, the out-of-plane dielectric constant, leakage current, and EOT of \ch{Mg(OH)_2} outperform bilayer h-BN. We verify the applicability of Anderson's rule, and conclude that bilayers of \ch{Ca(OH)_{2}} and \ch{Mg(OH)_{2}} respectively paired with lattice-matched monolayer \ch{HfS_{2}} and \ch{WS_{2}} are effective structural combinations that could lead to the development of innovative multi-functional Field Effect Transistors (FETs).

\end{abstract}
\section{Introduction}

The isolation of graphene has generated immense attention due to its unique physical properties such as efficient heat transport, exceptional optical nature (only absorbs 2.3\% of light over a wide range of frequencies), ballistic conductance, and unprecedented mechanical strength \cite{geim2010rise,novoselov2012roadmap, neto2009electronic, vandenberghe}. These properties have motivated many scientists to devote their efforts to exploring other two-dimensional (2D) van der Waals (vdW) materials. Different experimental techniques such as mechanical exfoliation \cite{gao2018mechanical}, epitaxial growth \cite{li2017epitaxial}, and chemical vapor deposition \cite{zhang2013review, cai2018chemical} have been employed to synthesize other prominent 2D vdW compounds, such as large and narrow bandgap semiconductors: hexagonal boron nitride (h-BN)\cite{zhang2017two}, and black phosphorus \cite{carvalho2016phosphorene,khandelwal2017phosphorene}. The continuous search for other promising 2D materials led to the discovery of the most recognized family of 2D layered materials, transitional metal dichalcogenides (TMDs), and prompted investigations into their unique and exceptional electronic, optical, and magnetic characteristics \cite{chhowalla2013chemistry,chhowalla2015two, schmidt2015electronic, choi2017recent, liu2021new, jalouli2021synthesis, jalouli2020spatial,reyntjens2020magnetic}.

\paragraph{}
Hitherto, most of the 2D vdW materials have been used as channel materials in electronic applications in conjunction with three-dimensional (3D) dielectric materials to create high-performance metal-oxide semiconductor field effect transistors (MOSFETs) \cite{schwierz2015two,roy2014field,lanza2022redefining}. According to Moore's Law, increasing circuit complexity through scaling alone is insufficient \cite{moore1965cramming,schaller1997moore}. Instead, the electrical performance of scaled MOSFETs must be enhanced by incorporating low-dimensional materials such as 2D layered dielectrics \cite{schwierz2015two,frank1998generalized, illarionov2020insulators}. Although traditional non-vdW dielectrics, such as \ch{SiO_{2}} and \ch{HfO_{2}}, provide high-k solutions for silicon-based semiconductor technologies \cite{li2019uniform,wallace2002alternative, wilk2001high}, they cannot be scaled when grown on 2D channel materials \cite{cao20182, huyghebaert20182d}. Moreover, the oxide dangling bonds at the interface between a 2D vdW material and the 3D oxides cause an inevitable charge exchange between the oxide and channel \cite{liu2019van,jariwala2017mixed}. The electric fields resulting from the dangling bonds and the defects introduced in the 2D vdW material will inevitably increase the scattering rate and lower the mobility of carriers in the channel \cite{liu2019van,su2021layered}.
\paragraph{}

Recently, the scientific community has begun to examine alternate 2D vdW dielectrics to address the problem of unpassivated bonds at the surface observed in \ch{HfS_{2}}. 2D vdW dielectrics promise a dielectric that is layered and has naturally passivated bonds \cite{osanloo2021identification,osanloo2022transition}. Integrating vdW channel materials with 2D vdW dielectrics will enable low-EOT low defect dielectrics on top of monolayer channels to reach the transistor's ultimate scaling limit. Unfortunately, only one van der Waals gate dielectric is currently available: h-BN \cite{khalid2020dielectric, laturia2018dielectric}. But, h-BN has an unacceptable leakage current if used in Complementary metal-oxide-semiconductor (CMOS) transistors and its small dielectric constant lead to an unacceptably low capacitive coupling for thicker h-BN layers \cite{osanloo2022transition}. Therefore it is imperative to investigate new classes of layered dielectrics to harness all the possibilities for making a new generation of miniaturized and high-performance transistors. 
\paragraph{}

In our recent works \cite{osanloo2021identification, osanloo2022transition}, we have employed Density Functional Theory (DFT) to investigate novel 2D vdW dielectrics which aim to identify new vdW dielectrics. In the first work, we introduced six new candidate materials, namely HoOI, LaOBr, LaOCl, LaOI, \ch{SrI_{2}}, and YOBr as potential candidates for \textit{n}-MOS and \textit{p}-MOS technologies \cite{osanloo2021identification}. In a second and more recent work \cite{osanloo2022transition}, we investigate the dielectric performance of a promising class of Transitional Metal Nitride Halides (TMNHs) to examine their potential application in \textit{p}-MOS technology when combined with TMD channels. In this second work, a \ch{MoSe_{2}} channel with HfNCl as a gate dielectric is predicted to be the best combination for a \textit{p}-MOS transistor. Furthermore, there have also been experimental efforts attempting to resolve the issue, as recently reported in Nature Electronics regarding \ch{CaF_{2}} and \ch{Bi_{2}SeO_{5}} \cite{rao1966dielectric, wen2020dielectric, li2020native}. \ch{CaF_{2}} is a crystalline compound that can potentially address the low dielectric constant ($\sim$ 3.9), excessive leakage current, premature dielectric breakdown, and synthesis high temperature requirement in h-BN \cite{zhang2019thickness}. Moreover, \ch{CaF_{2}} eliminates some of the drawbacks of an amorphous oxide like \ch{HfO_{2}} and \ch{SiO_{2}}. However \ch{CaF_{2}} does not alleviate the issue of unpassivated bonds at the surface. \ch{Bi_{2}SeO_{5}} is a layered material that was grown as a native oxide on the layered \ch{Bi_{2}O_{2}Se} \cite{li2020native}, but it is still unclear how it would be grown on other materials. Consequently, more research needs to be conducted and delicate experimental methods should be employed to measure and verify the actual performance of the 2D vdW materials and their optical and dielectric properties \cite{xu2021optical, dell2019optical, dell2022reflection}.

\paragraph{}

Calcium hydroxide, \ch{Ca(OH)_{2}}, and magnesium hydroxide, \ch{Mg(OH)_{2}} are prominent members of a class of multi-functional 2D layered inorganic compounds with a wide range of applications in cutting-edge technologies in electronics and photo-electronic devices \cite{lou2021electronic, luo2019transition, ozcelik2018highly, xia2017electric}. They are the simplified examples of isomorphous hydroxides with a chemical formula of \ch{M(OH)_{2}}, where M (= Ca, Mg) is a alkali-earth metal and (OH) is known as hydroxide. The tightly bonded hydrogen and oxygen atoms in -OH groups form chemically passivated surfaces, which explains the stability of these 2D structures under ambient conditions. The molecules in both compounds are held together via ionic bonds between the calcium ion \ch{(Ca$^{2+}$)} and two hydroxide ions \ch{(OH$^{–}$)} \cite{baranek2001structural,jochym2010structure, xia2017robust}. While in our previous study \cite{osanloo2021identification}, we dismissed \ch{Ca(OH)_{2}} and \ch{Mg(OH)_{2}} because they are elemental bases and soluble in water, they are not very soluble and their solubility decreases significantly with temperature\cite{pannach2017solubility} making these compounds suitable for variety of industrial applications. Interestingly, large sample sizes of \ch{Ca(OH)_{2}} ($>$5 mm), and \ch{Mg(OH)_{2}} ($>$8 mm) were recently grown (by 2D semiconductors USA) \cite{2dsemiconductors} using the float zone synthesis technique to yield perfectly layered and highly crystalline vdW crystals.

\ch{Ca(OH)_{2}}, is the only known hydroxide of calcium. The mineral form of \ch{Ca(OH)_{2}} is sometimes called portlandite. It is well-established that crystallized \ch{Ca(OH)_{2}} in dry air is stable and has an easily cleavable layered brucite type structure \cite{lieth1977preparation}. \ch{Mg(OH)_{2}} is also found in a mineral form known as abrucite or texalite and can be synthesized using different techniques \cite{lieth1977preparation,meyer1926gmelins}. In addition to their dielectric properties, \ch{Ca(OH)_{2}} and \ch{Mg(OH)_{2}} are examples of advanced materials with applications in carbon capture and heat storage \cite{zhang2021simultaneous}. Therefore, these materials are available for experimental studies and opportunities for their application in cutting-edge technologies are unexplored. 

\paragraph{}

In this work, we perform accurate first-principles calculations to study the electronic, thermodynamic, and dielectric properties of two novel vdW materials: \ch{Ca(OH)_{2}}, and \ch{Mg(OH)_{2}}. In addition to the electronic properties of monolayer \ch{Ca(OH)_{2}} and \ch{Mg(OH)_{2}}, we report the exfoliation energies of these two promising layered dielectrics along with their thermodynamic stability from their phonon spectrum. We also study their thermodynamic properties including the free energy, their heat capacitance at constant volume, and entropy change at various temperatures. Moreover, we accurately calculate the macroscopic in-plane and out-of-plane dielectric constants of the bulk and monolayer using Density Functional Perturbation Theory (DFPT). We also use the HSE06 hybrid functional to calculate the bandgaps, electron affinities, and the effective masses of charge carriers in the monolayer. We model the performance of each of these materials as a gate dielectric, considering its equivalent oxide thickness (EOT) as well as its leakage current. We consider the performance of monolayer \ch{Ca(OH)_{2}} and \ch{Mg(OH)_{2}} as dielectrics when combined with monolayer \ch{HfS_{2}} and \ch{WS_{2}} channels. Although a stringent lattice matching requirement is not required for 2D heterostructures, the existence of substantial lattice mismatch combined with weak vdW bonding between the 2D layers can result in incoherent lattice matching and the formation of Moiré patterns \cite{kang2013electronic}. Therefore, to design closely aligned heterostructure with minimum lattice mismatch ($<$ 0.5\%), we integrate \ch{Ca(OH)_{2}} with \ch{HfS_{2}} and \ch{Mg(OH)_{2}} with \ch{WS_{2}}.

\section{Results and Discussion}

\begin{figure}[ht]
    \centering
    \includegraphics[width=1\textwidth]{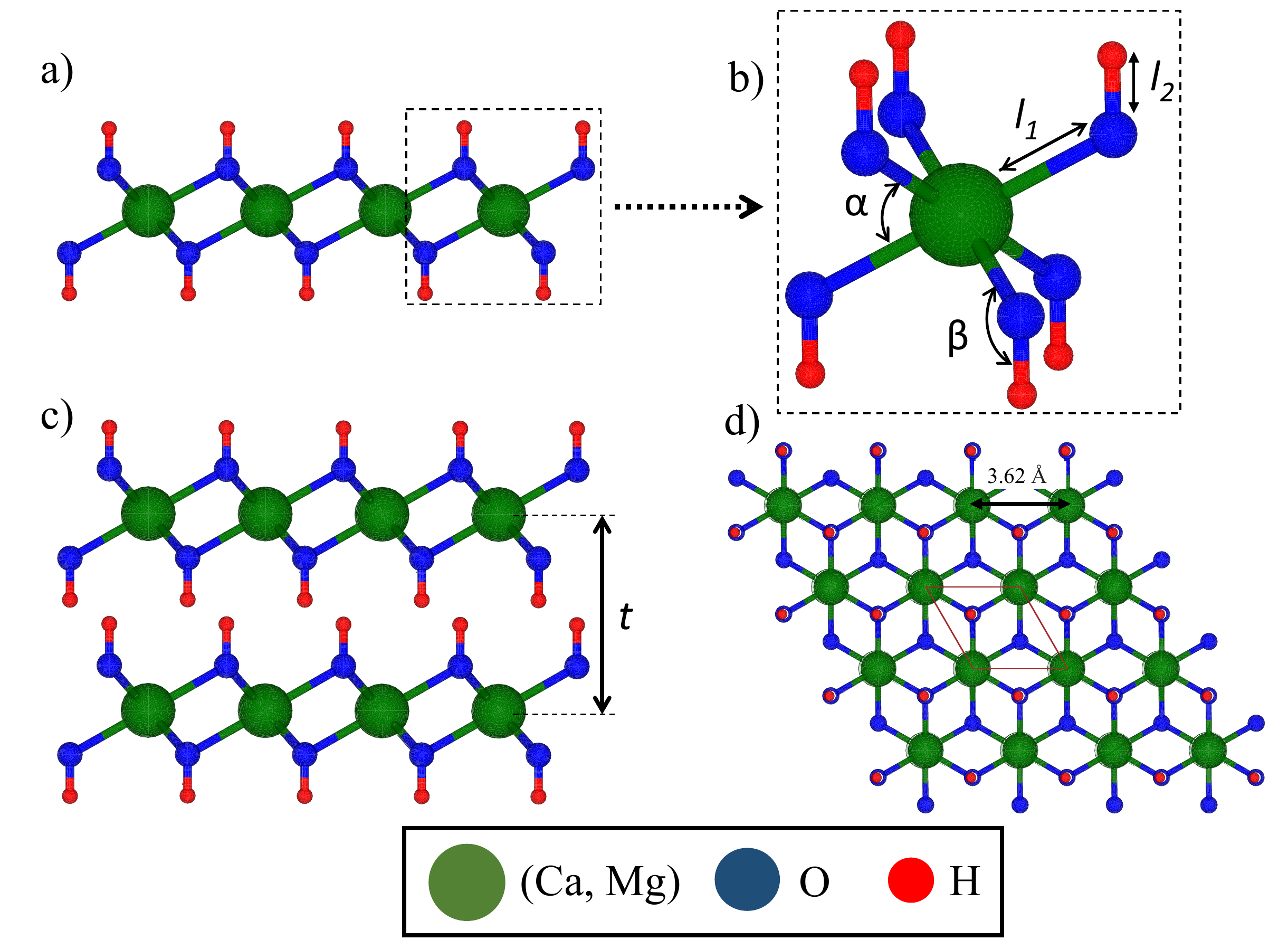}
    \caption{Structures of alkaline-earth metal hydroxides. The side view (a) and top view of the monolayer(d) in addition to the side view of the bilayer (c) are illustrated. (b) demonstrates  Ca-O/Mg-O ($l_{1}$) and O-H bond length ($l_{2}$) along with the angles between bonds ($\alpha$ and $\beta$). The monolayer thickness ($t$) is indicated on the bilayer structures.}
    \label{fig:structure}
\end{figure}

Fig~\ref{fig:structure} depicts side and top views of the atomic structure of the monolayer metal hydroxides under investigation. The layered crystal structure of \ch{M(OH)_{2}} has a hexagonal shape with an A-A stacking configuration in which the hydroxides form a relatively close-packed array, marginally expanded in the out-of-plane direction, surrounding cations $(X)^{2+}$ in octahedral coordination. Table~\ref{tab:parameters} reports the structural parameters such as monolayer thickness (\emph{t}), bulk interlayer distance (\emph{d}), and the lattice constant of these metal hydroxides.  The monolayer thicknesses (obtained from bilayer) of \ch{Ca(OH)_{2}} and \ch{Mg(OH)_{2}}  are 4.78 Å and 4.60 Å, whereas the calculated interlayer distance (obtained from bulk) of these compounds is 4.75 Å and 4.56 Å, respectively. Sequentially, the optimized in-plane lattice constants of monolayer \ch{Ca(OH)_{2}} and \ch{Mg(OH)_{2}} are calculated to be 3.61 Å and 3.28 Å. As depicted in Fig~\ref{fig:structure}, the bond lengths of Ca-O (Mg-O) and O-H bonds are calculated to be $l_{1}$=2.38 Å (2.15 Å) and 0.97 Å, respectively. Furthermore, the calculated angles of O-Ca-O (O-Mg-O) and Ca-O-H (Mg-O-H) in the optimized structures are $\alpha$=81.13$^{\circ}$ (80.34$^{\circ}$) and $\beta$=118.69$^{\circ}$ (118.07$^{\circ}$), respectively. The calculated lattice parameters and bond lengths calculated in this study are in consonance with the values reported in previous experimental and theoretical works \cite{yagmurcukardes2019raman,baranek2001structural,yagmurcukardes2016mg,busing1957neutron}.

We compute the exfoliation energies, \ch{E_{ex}}, to verify if \ch{Ca(OH)_{2}} and \ch{Mg(OH)_{2}} are indeed layered. The calculated exfoliation energy values for \ch{Ca(OH)_{2}} and \ch{Mg(OH)_{2}} are 33.48 meV/$\text{\AA}^{2}$ and 57.91 meV/$\text{\AA}^{2}$, respectively. Taking the general-guideline in Ref. \cite{duong2017van} stating that exfoliable compounds have \ch{E_{ex}} $<$ 100 meV/$\text{\AA}^{2}$, both \ch{Ca(OH)_{2}} and \ch{Mg(OH)_{2}} are easily exfoliable, albeit not as easily exfoliable as TMD materials which have a \ch{E_{ex}} $<$ 25 meV/$\text{\AA}^{2}$. Experimentally, monolayer \ch{Ca(OH)_{2}} and \ch{Mg(OH)_{2}} have been exfoliated from the bulk portlandite crystal and from its synthesized bulk crystals onto piranha cleaned 285 nm thermal \ch{SiO_{2}}/Si substrates \cite{suslu2016unusual, ozcelik2018highly}.

\begin{table}
	\caption[Structural parameters of \ch{M(OH)_{2}}]{Structural parameters, monolayer thickness, interlayer distance, and exfoliation energies of \ch{Ca(OH)_{2}} and \ch{Mg(OH)_{2}}. Exfoliation energies $<$ 100 meV/$\text{\AA}^{2}$ indicate easily exfoliable materials \cite{duong2017van}.}
    \includegraphics[width=1\textwidth ]{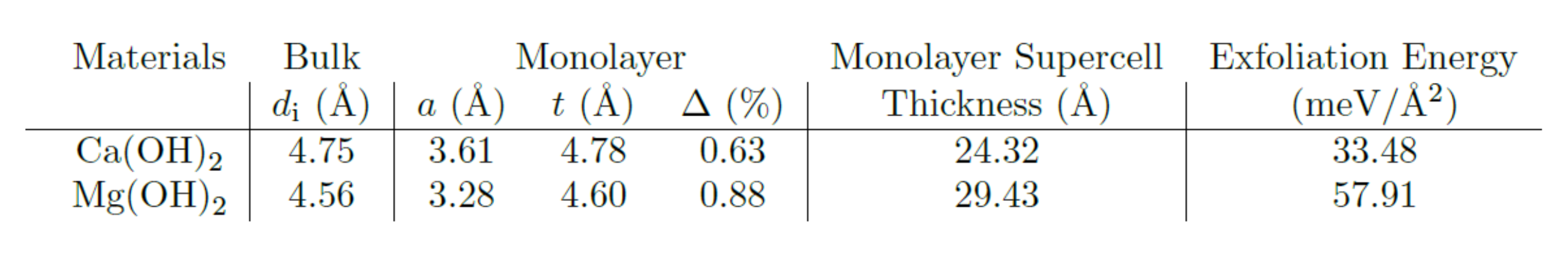}
    \label{tab:parameters}
\end{table}

\begin{figure}[ht]
    \centering
    \includegraphics[width=1\textwidth ]{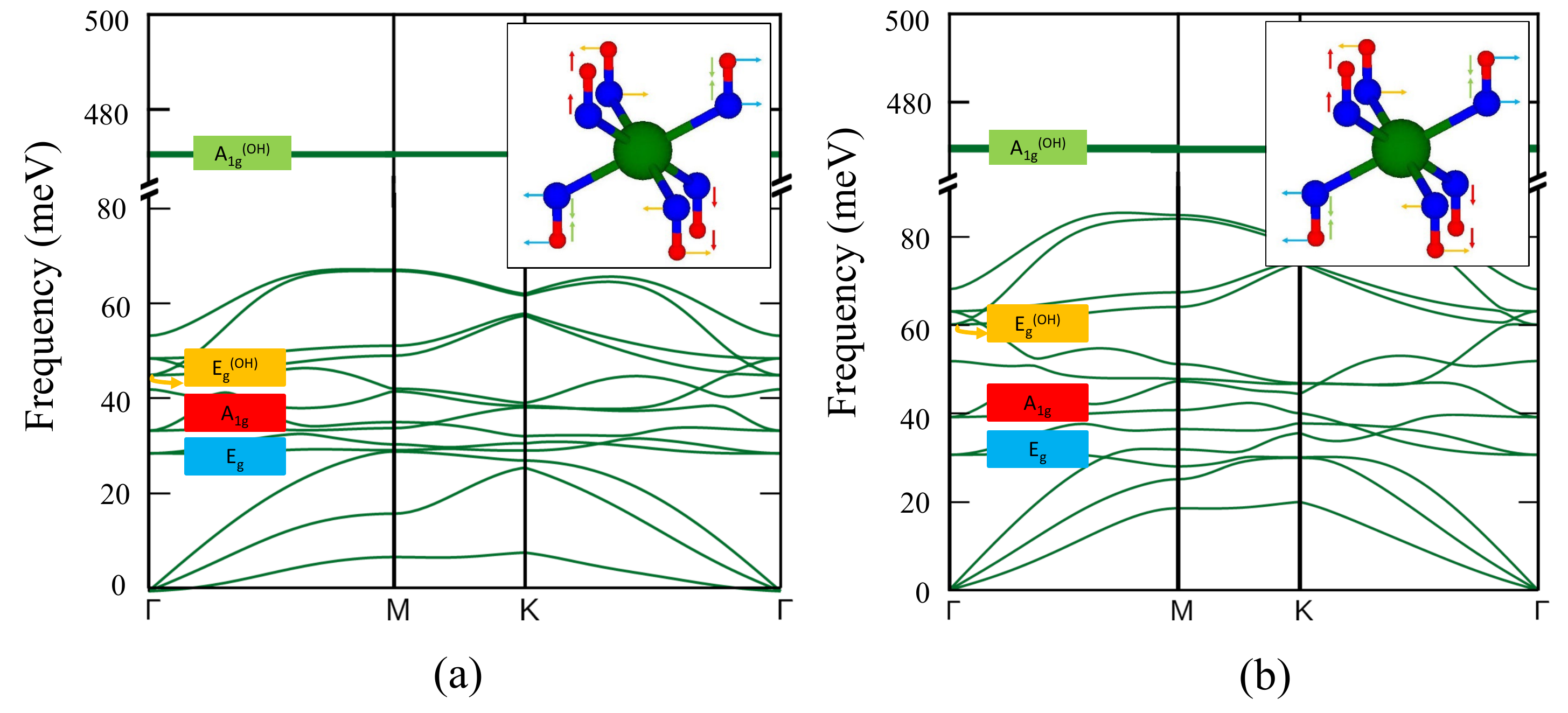}
    \caption{\ch{M(OH)_{2}}'s phonon dispersion curves. (a) Phonon dispersion spectrum of \ch{Ca(OH)_{2}}. (b) Phonon dispersion spectrum of \ch{Mg(OH)_{2}}. {\ch{E_{g}}, \ch{A_{1g}}} and {\ch{E_{g}^{(OH)}}, \ch{A_{1g}^{(OH)}}} modes respectively represent reflect the translational motion, and the reciprocating motion of the O-H bonds. The broken axis represents that there is no phonon branches between 90 meV and 450 meV. The flat energy curves at high-energy modes ($\sim$ 470 meV) are associated with out-of-plane hydrogen and oxygen displacements.} 
    \label{fig:phonons}
\end{figure}

Fig~\ref{fig:phonons} shows the 2D phonon dispersion curves for monolayer \ch{Ca(OH)_{2}} (a) and \ch{Mg(OH)_{2}} (b). Since the unit cell of monolayer \ch{Ca(OH)_{2}} and \ch{Mg(OH)_{2}} includes five atoms, the phonon dispersion has twelve optical and three acoustic modes. The phonon spectra, in Fig~\ref{fig:phonons}, with strictly positive frequencies clearly indicate that the monolayer of both materials is predicted to be thermodynamically stable. \ch{E_{g}} and \ch{A_{1g}} modes reflect translational motion, implying that the O-H bond distance is usually constant, but \ch{E_{g}^{(OH)}} and \ch{A_{1g}^{(OH)}} modes represent reciprocating vibration of O and H atoms, implying that the O-H bond distance varies. In both compounds, a high frequency mode with energy 470 meV appears in the phonon spectrum. Inspecting the displacement vectors of this high-frequency mode ($\sim$ 470 meV) we see that it is associated with out-of-plane displacement of the hydrogen atoms.

\begin{figure}[ht]
    \centering
    \includegraphics[width=1\textwidth ]{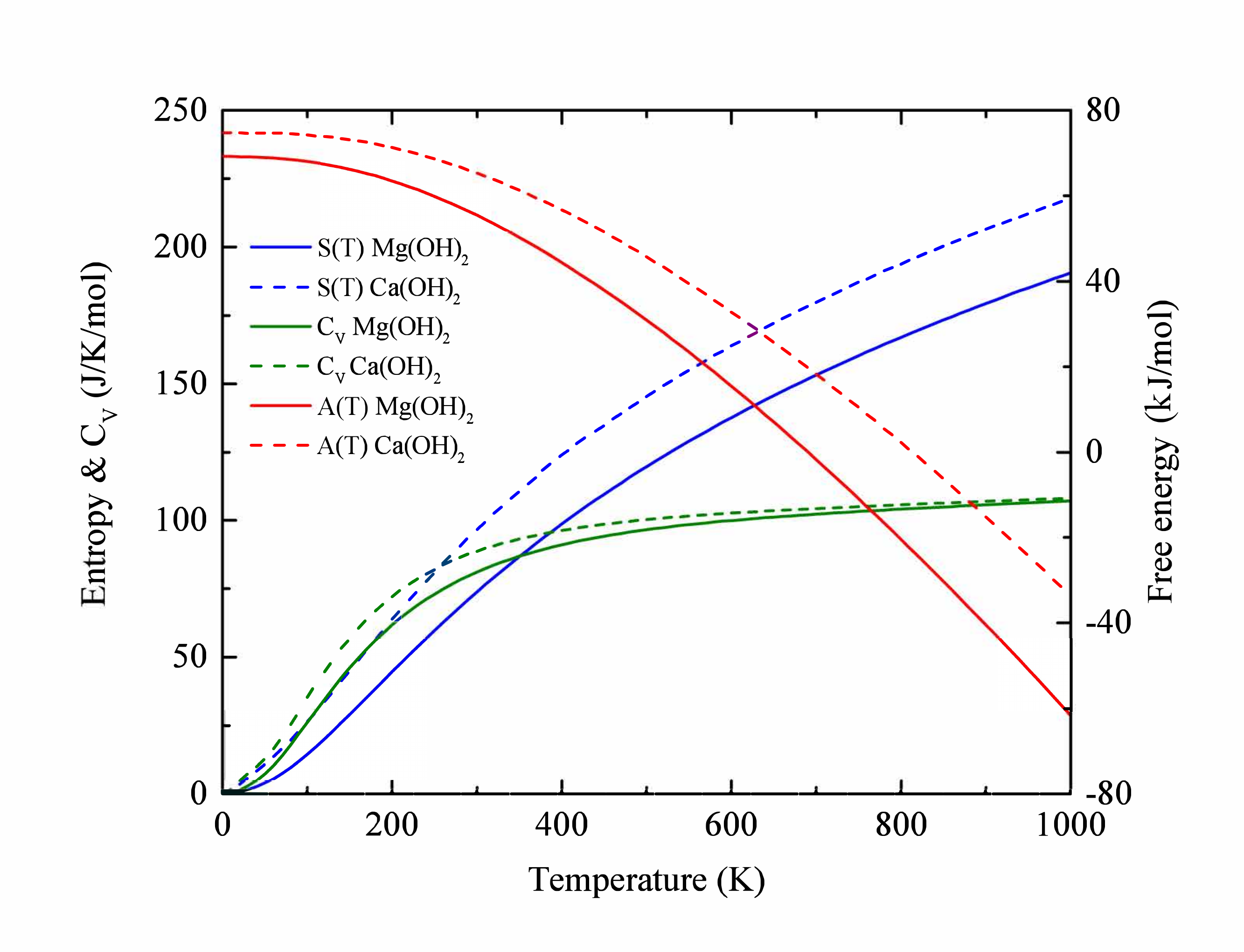}
    \caption{Thermodynamic properties of  \ch{Ca(OH)_{2}} and  \ch{Mg(OH)_{2}}. Entropy($S_{v}(T)$) , constant volume heat capacity ($C_{v}$), and Helmholtz Free energy (A(T)) of \ch{Mg(OH)_{2}} (\ch{Ca(OH)_{2}}) are shown with solid (dashed) blue, green, and red lines, respectively. The smaller Helmholtz free energy implies that from a thermodynamic prospective, \ch{Mg(OH)_{2}} is the better material compared to \ch{Ca(OH)_{2}.}}
    \label{fig:thermal}
\end{figure}

Fig~\ref{fig:thermal} shows the thermodynamic properties of monolayer \ch{Ca(OH)_{2}} and \ch{Mg(OH)_{2}}, in which $S_{v}(T)$, $C_{v}$, and A(T), are the vibrational entropy, the heat capacity at constant volume, and the Helmholtz free energy of a 2D system (see computational methods). Our calculation reveals a steeper change in the entropy for \ch{Ca(OH)_{2}} which is an indicator of a higher rate of chemical reactivity. We observe that $C_{v}$ of monolayer \ch{Ca(OH)_{2}} at around room temperature is slightly higher than that of monolayer \ch{Mg(OH)_{2}}, specifying a greater change in the internal energy in a wide range of temperatures. The Helmholtz free energy in \ch{Mg(OH)_{2}} is about 30 kJ/mol smaller than the Helmholtz free energy of \ch{Ca(OH)_{2}} predicting that the surface structure of \ch{Mg(OH)_{2}} over a wide temperature range (from 0 to 1000 K) is more thermodynamically favorable. So from a thermodynamic prospective, \ch{Mg(OH)_{2}} is the better material compared to \ch{Ca(OH)_{2}}. 
 
\paragraph{}

\begin{figure}[ht]
    \centering
    \includegraphics[width=1\textwidth ]{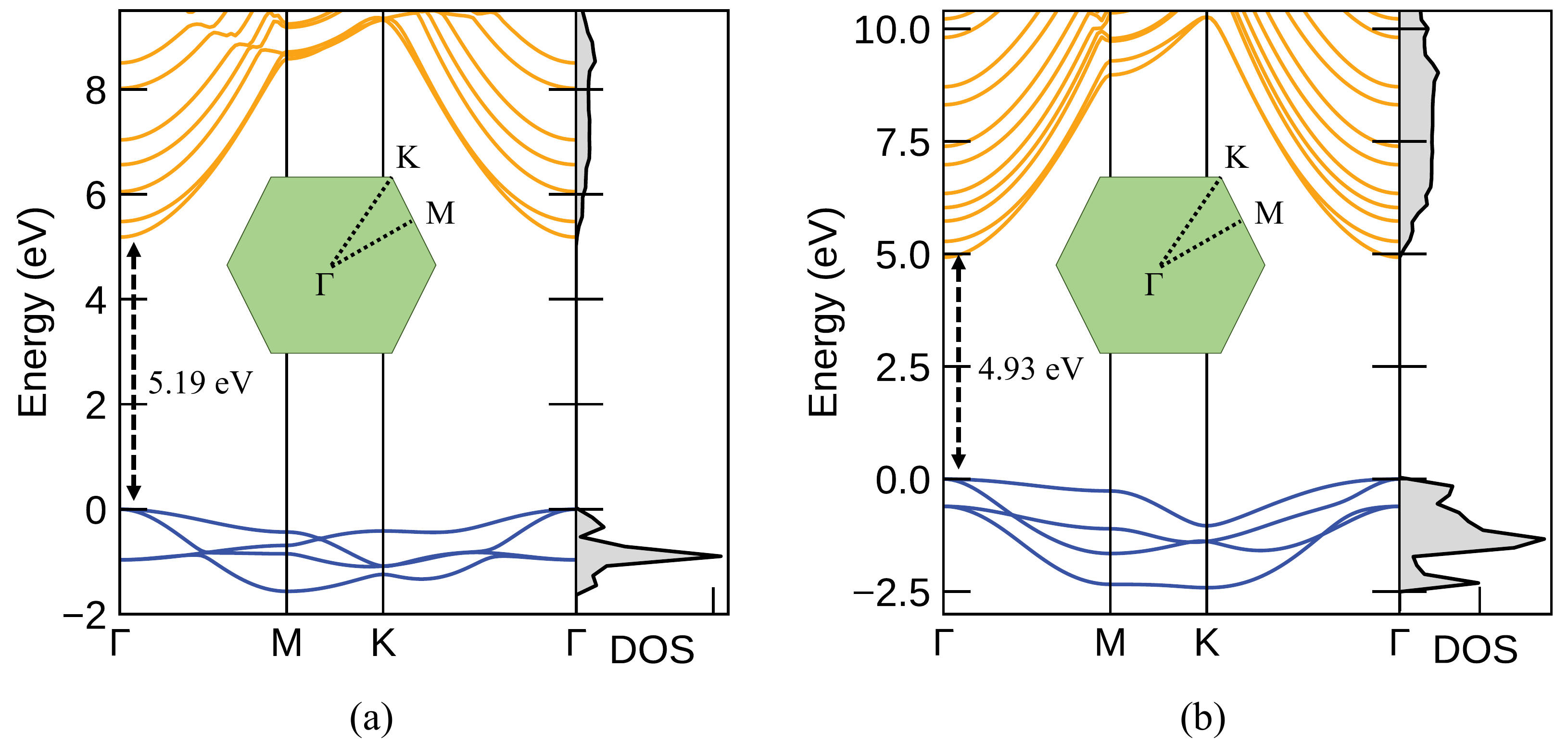}
    \caption{The band structure and total DOS of (a) monolayer \ch{Ca(OH)_{2}} (b) \ch{Mg(OH)_{2}}. HSE06 hybrid functionals are used to correct for the usual bandgap underestimation. Both materials have a direct bandgap with the \ch{Ca(OH)_{2}} bandgap (5.19 eV) slightly larger than the \ch{Mg(OH)_{2}} bandgap (4.93 eV).}
    \label{fig:band_dos}
\end{figure}

Fig~\ref{fig:band_dos} illustrates the band structures and total DOS of monolayer \ch{Ca(OH)_{2}} and \ch{Mg(OH)_{2}}  along the high symmetry k-points ($\Gamma$-M-K-$\Gamma$). We calculated the bandgap of these materials using PBE and HSE06. The Heyd–Scuseria–Ernzerhof hybrid functional (HSE06) compensates for the bandgap underestimation observed in non-hybrid PBE calculations. The calculated bandgaps from PBE for the monolayer \ch{Ca(OH)_{2}} and \ch{Mg(OH)_{2}} are 3.68 eV and 3.42 eV, respectively, whereas the monolayer bandgaps calculated using HSE06 are 5.19 eV and 4.93 eV. As we see from the band structure, both monolayer compounds have a direct bandgap with valence and conduction bands located at the $\Gamma$ point. Our calculation shows that both PBE/HSE06 predict a slightly larger bandgap for monolayer \ch{Ca(OH)_{2}} compared to \ch{Mg(OH)_{2}}.

To explore the insulating properties of \ch{Ca(OH)_{2}} and \ch{Mg(OH)_{2}} monolayer dielectrics, we compute the band offset of monolayer \ch{Ca(OH)_{2}} and \ch{Mg(OH)_{2}} and compare it with two different TMD channels: \ch{HfS_{2}}, and \ch{WS_{2}}. In addition to exfoliability and stability as  characteristics of a good vdW layered dielectric, a desirable dielectric must be a good insulator with a suitable dielectric-channel band offset surpassing at least 1 eV to reduce leakage current through tunneling or thermionic emission.

\begin{figure}[ht]
    \centering
    \includegraphics[width=1\textwidth ]{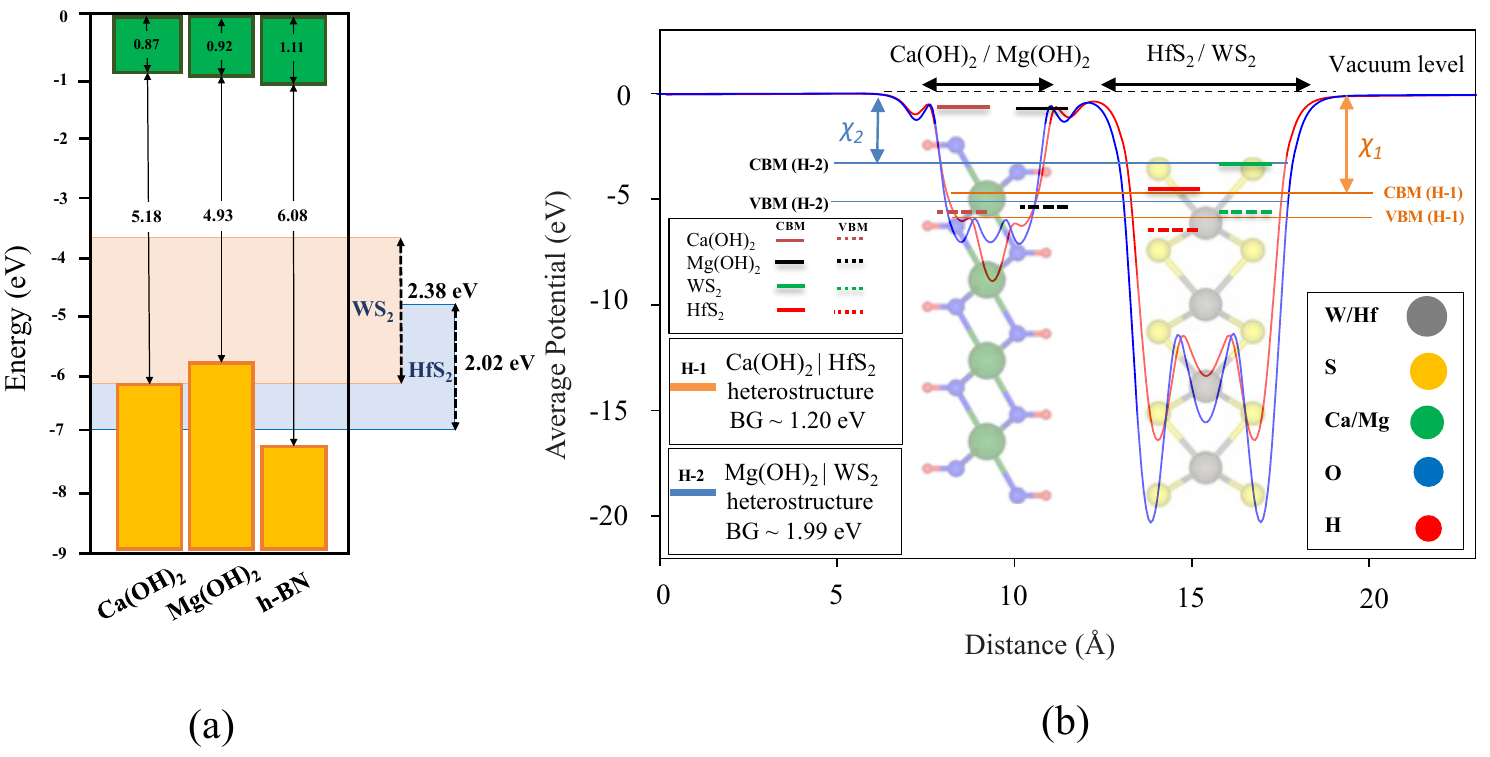}
    \caption{The band offset for single-layer \ch{Ca(OH)_{2}} and \ch{Mg(OH)_{2}} are shown. The vacuum level is set to zero and a monolayer of h-BN is included for the purpose of comparison. The average potential of the heterostructures (\ch{Ca(OH)_{2}}/\ch{HfS_{2}} and \ch{Mg(OH)_{2}}/\ch{WS_{2}}) is shown with respect to the distance change along the $z$-direction, perpendicular to the plane of sheets. Anderson's rule works well for \ch{Ca(OH)_{2}}/\ch{HfS_{2}} heterostructure.}
    \label{fig:bandoffset}
\end{figure}

Fig\ref{fig:bandoffset}-a displays the electron affinity, the bandgap, and the relative position of the band edges (with respect to the vacuum level) of monolayer \ch{Ca(OH)_{2}} and \ch{Mg(OH)_{2}} with \ch{HfS_{2}} and \ch{WS_{2}} as channel materials. The conduction band and the valence band edges are indicated by the solid green and yellow lines, respectively. As shown in Fig~\ref{fig:bandoffset}-a, the 1 eV band offset requirement of each dielectric with the valence band of the proposed TMD channels (\ch{HfS_{2}} and \ch{WS_{2}}) is not met, signifying that only designing an \emph{n}-MOS transistor with \ch{Ca(OH)_{2}}/\ch{HfS_{2}} and \ch{Mg(OH)_{2}}/\ch{WS_{2}} is theoretically feasible. Therefore, we only evaluate the performance of these dielectrics in an $n$-MOS transistor. Fig. 5-b displays the averaged potential of the heterostructures (\ch{Ca(OH)_{2}}/\ch{HfS_{2}} and \ch{Mg(OH)_{2}}/\ch{WS_{2}}) with respect to the distance along the z-axis of the supercell lattice. Applying Anderson's rule, the HSE bandgaps for \ch{Ca(OH)_{2}}/\ch{HfS_{2}} and \ch{Mg(OH)_{2}}/\ch{WS_{2}} are calculated to be 1.20 eV and 1.99 eV, while we calculate 0.97 eV, and 2.11 eV band offset for monolayer \ch{Ca(OH)_{2}} and \ch{Mg(OH)_{2}} when combined with \ch{HfS_2} and \ch{WS_2}, respectively. As depicted, the electron affinity($\chi$), is the difference between the vacuum level and the Fermi level (vacuum level is shifted to zero). The electron affinity of \ch{Ca(OH)_{2}}/\ch{HfS_{2}} and \ch{Mg(OH)_{2}}/\ch{WS_{2}} obtained from HSE06 are calculated to be 3.12 eV and 3.00 eV, respectively. We also include the band edges of each \ch{M(OH)_{2}} and TMD compare with Anderson's rule \cite{anderson1960germanium} for heterostructures. As depicted in Fig~\ref{fig:bandoffset}-b, the predicted HSE bandgap using Anderson's rule for the \ch{Ca(OH)_{2}}/\ch{Hf_{2}} and \ch{Mg(OH)_{2}}/\ch{WS_{2}} heterostructures are in good agreement with the calculated HSE bandgap for \ch{Mg(OH)_{2}} \ch{WS_2} channel.

Table~\ref{tab:dielectric_constants} shows the calculated bulk dielectric constants in the in-plane ($\parallel$) and out-of-plane ($\perp$) directions. In order, the calculated in-plane static dielectric constants of \ch{Ca(OH)_{2}} and \ch{Mg(OH)_{2}} are 12.30 and 9.75 while in the out-of-plane direction, the calculated values are 4.53 and 4.32, respectively. In both materials, the optical dielectric constant is much lower in the in-plane direction than in the out-of-plane direction, similar to what we found in previous studies on layered materials \cite{laturia2018dielectric,osanloo2021identification}. We find that in the bulk materials, the ionic contribution to the static dielectric constant is large and accounts for 335\% (260\%) for the in-plane and for the 75\% (62\%) out-of-plane dielectric response in \ch{Ca(OH)_{2}} (\ch{Mg(OH)_{2}}).

To compute the monolayer dielectric constant, we isolate monolayers in a computational supercell with sufficient vacuum, then we rescale the supercell's estimated dielectric values to the monolayer's, as detailed in the computational methods. Our calculations demonstrate that monolayer of \ch{Ca(OH)_{2}} and \ch{Mg(OH)_{2}} have static in-plane dielectric constants of 8.94 and 7.75, whereas their out-of-plane static dielectric constants are 6.40 and 6.33, respectively. Our results show a decrease in the in-plane dielectric constant when we move from bulk to monolayer, while the trend is increasing in the out-of-plane direction.

\begin{table}
	\caption[Monolayer and bulk dielectric constants of \ch{Ca(OH)_{2}} and \ch{Mg(OH)_{2}}.]{The static dielectric constants of bulk and monolayer \ch{Ca(OH)_{2}} and \ch{Mg(OH)_{2}}. The in-plane dielectric monolayer \ch{Ca(OH)_{2}} is higher constant, while in the out-of-plane direction, \ch{Mg(OH)_{2}} has a higher dielectric constant.}
    \includegraphics[width=1\textwidth ]{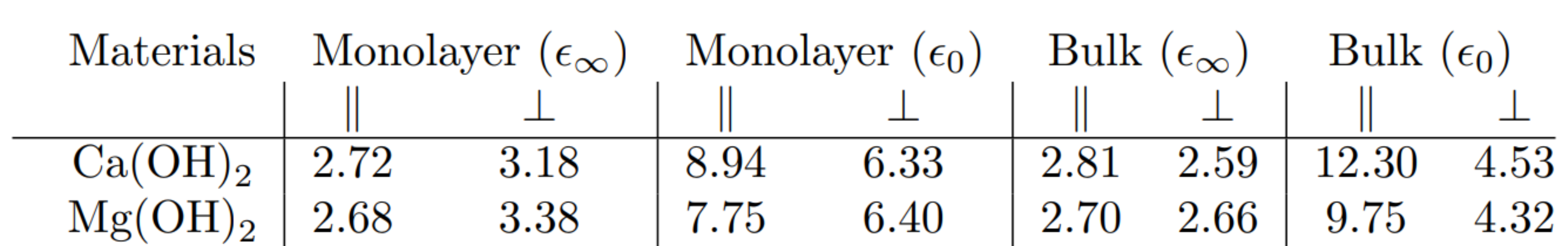}
    \label{tab:dielectric_constants}
\end{table}

Compared to the dielectric constants of wide bandgap ionic crystals such as \ch{ZnCl_{2}} (4.00), CaHBr (4.60), and \ch{MgF_{2}} (5.40) \cite{osanloo2021identification, babu2011structural}; monolayer \ch{Ca(OH)_{2}} offers a higher out-of-plane static dielectric constant. Moreover, the dielectric constants of a monolayer \ch{Ca(OH)_{2}} and \ch{Mg(OH)_{2}} are greater than the dielectric constant of monolayer h-BN (3.29) \cite{osanloo2021identification, laturia2018dielectric}. Moreover, while \ch{Ca(OH)_{2}} and \ch{Ca(OH)_{2}} do not match the dielectric constants of "high-k" materials like \ch{HfO_{2}}, it is significantly higher than \ch{SiO_{2}} or h-BN. \cite{wallace2002alternative, wallace2003high}.

\subsection{Tunneling Current and Dielectric Performance}
To design a viable transistor, a dielectric with a small thickness, a high dielectric constant, and a low leakage current is desirable. As outlined in the computational methods, to quantify the promise of a gate dielectric material we compute the leakage current accounting for direct tunneling and thermionic emission for low power devices.

\paragraph{}

To find the best dielectrics, we compute the Equivalent Oxide Thickness (EOT) of  \ch{Ca(OH)_{2}} and \ch{Mg(OH)_{2}} and leakage current, assuming the channel materials  (\ch{HfS_{2}} and \ch{WS_{2}}), in an $n$-MOSFET with an electron affinity of 4.98 eV and 3.73 eV, respectively.  In addition to the stability and lower leakage current criteria, a suitable dielectric candidate should have a small EOT to ensure acceptable electrostatic control. As a reference and for a better comparison of the performance of a device, the leakage current of monolayer h-BN is also calculated. According to the International Roadmap for Devices and Systems (IRDS) \cite{hoefflinger2020irds}, the absolute maximum leakage current for any feasible gate dielectric is less than 100 pA/$\mu$m per pitch for a transistor with a 28 nm pitch, an effective gate width of 107 nm, and a 18 nm long gate. With these criteria, the acceptable current density is about 0.145 A/$\ch{cm^{2}}$. We remark that we used a larger k-grid (15 $\times$ 15 $times$ 1) for the HSE calculations in this study, , improving over previous estimations \cite{osanloo2021identification, osanloo2022transition, laturia2018dielectric}.

Table~\ref{tab:leakage_current} shows the leakage currents of \ch{Ca(OH)_{2}} and \ch{Mg(OH)_{2}} and the calculated EOTs. Our calculations explicitly show that a monolayer of \ch{Mg(OH)_{2}} with an EOT $\leq$ 0.3 nm satisfies the minimum leakage current criteria as determined by IRDS, whereas \ch{Ca(OH)_{2}} when combined with \ch{HfS_{2}} does not adequately block leakage current. In comparison, a very small physical thickness of monolayer h-BN and its small dielectric constant result in high leakage currents, making monolayer h-BN unfit for use as a gate insulator in 2D transistors \cite{knobloch2021performance}. We added the leakage current of bilayers for the purpose of comparison. Although monolayer h-BN is not sufficiently insulating for $n$-MOS, bilayer h-BN with a higher physical thickness has a smaller leakage current($<$ 6.37$\times$$10^{-13}$ A/\ch{cm^{2}}) acceptable by IRDS. According to Table~\ref{tab:leakage_current}, a monolayer of \ch{Ca(OH)_{2}} with a \ch{HfS_{2}} channel is not sufficiently insulating, however, bilayer of \ch{Ca(OH)_{2}} still has a low EOT ($\sim$ 0.56 nm) while small leakage current is small ($<$ 6.06$\times$$10^{-8}$ A/\ch{cm^{2}}). Finally, taking the result from our previous study, monolayer and bilayer of LaOCl considerably outperform both monolayer and bilayer of \ch{M(OH)_{2}} and h-BN when combined with \ch{HfS_{2}} and \ch{WS_{2}} channels. We observe that the best performance of a dielectric/channel heterostructure is found for monolayer LaOCl/\ch{HfS_{2}} with a small EOT (0.05 nm) and leakage current (4.79$\times$$10^{-21}$ A/$\ch{cm^{2}}$). While alternative layered dielectrics, such as LaOCl, may give even better performance, \ch{Ca(OH)_2} and \ch{Mg(OH)_2} are both found in nature and can be commercially synthesized on sizeable crystals.

\begin{table}
	\caption[Out-Of-Plane Electron Effective Mass, Leakage Current, and EOT of 1L \ch{M(OH)_{2}}, h-BN, and LaOCl]{Leakage current density for $n$-MOS applications through a monolayer (1L), and bilayer (2L) \ch{Ca(OH)_{2}} and \ch{Mg(OH)_{2}} with \ch{HfS_{2}} and \ch{WS_{2}} as channel materials, respectively. For comparison, the leakage current of monolayer \& bilayer h-BN as well as LaOCl are included.}
    \includegraphics[width=1\textwidth ]{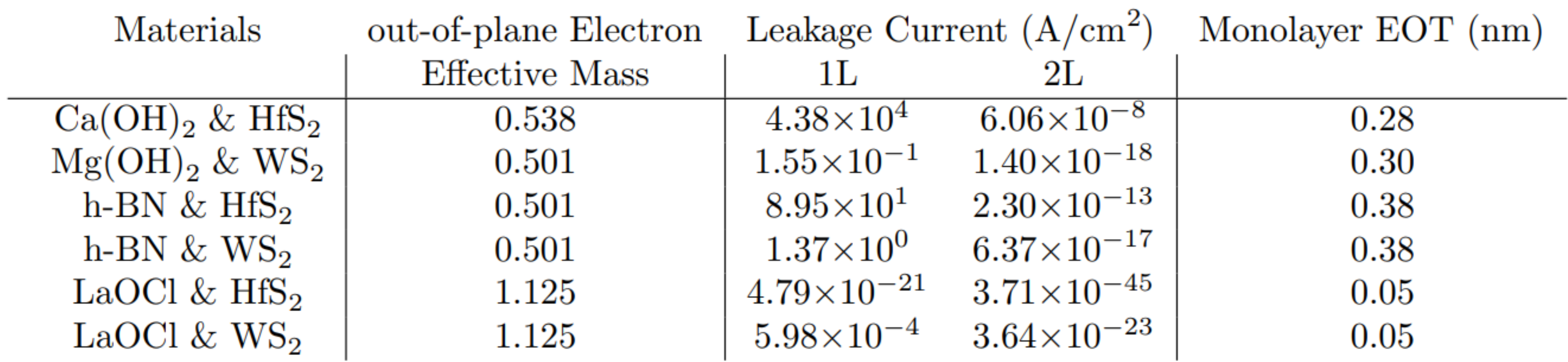}
    \label{tab:leakage_current}
\end{table}

\section{Computational Methods}
We employ DFT as implemented in the Vienna ab initio simulation package (VASP) \cite{heyd2004efficient} and use the generalized gradient approximation (GGA) as proposed by Pedrow-Burke-Ernzerof (PBE) for the electron exchange and correlation functional \cite{perdew1996generalized}. We set the plane-wave basis energy cut-off to 520 eV for monolayer and bulk \ch{Ca(OH)_{2}} and \ch{Mg(OH)_{2}}. The structural relaxations are continued until the force on each atom is less than $10^{-3}$ eV/Å. For precise phonon and dielectric calculations, we set a tight energy convergence criteria of $10^{-8}$ eV. We employ the same criteria for monolayer TMDs (\ch{HfS_{2}} and \ch{WS_{2}}) as we used in our prior work in Ref \cite{osanloo2022transition}. To mesh the Brillouin Zone (BZ), 12$\times12\times$12 and 12$\times12\times$1 k-points grids are employed for the bulk and the monolayer structures, respectively. For the heterostructures, we used 15$\times15\times$1 k-points grids for the structural relaxation and 6$\times6\times$1 for the HSE06 bandgap calculation. The DFT-D3 approach of Grimme is used to account for interlayer van der Waals interactions \cite{grimme2006semiempirical}. We establish at least 15 Å vacuum between the monolayers to avoid any non-physical interactions between layers. Moreover, the exfoliation energy is calculated as the ratio of the difference in bulk and monolayer ground state energies to the surface area of the bulks \cite{bjorkman2012van, ashton2017topology}. We used the sumo code to plot band structures and Density of States (DOS)\cite{ganose2018sumo}.

\paragraph{}

To calculate phonon spectrum we use the open-source package; Phonopy \cite{togo2015first}. The first-principles phonon calculations with a finite displacement method (FDM) are generated for a set of displacements \cite{vandenberghe}. We use a $4 \times 4 \times 1$ supercell yielding a total of 80 atoms where the atomic displacement distance is $10^{-3}$ Å. When the phonon frequencies over the Brillouin zone are computed, the Helmholtz free energy of the phonons under the harmonic approximation can be estimated using the canonical distribution in statistical mechanics, as detailed in Ref \cite{togo2015first,maradudin1963theory}. Once the phonon frequencies are calculated, we can use thermodynamic relations to determine the thermodynamic properties of the system:

\begin{equation}
    A(T) = U(T) - TS(T)
\end{equation}

\begin{equation}
    C_{v} = \left(\frac{\partial U}{\partial T}\right)_{V}
\end{equation}

\begin{equation}
    S = \frac{\partial A}{\partial T}
\end{equation}

where $A(T)$, $C_{v}$, and $S$ are the Helmholtz free energy, constant volume heat capacity, and entropy, respectively. $U(T)$ = $U_{L}$ + $U_{V}(T)$ is the phonon-contribution to the internal energy of the system which is a combination of lattice internal energy ($U_{L}$), and vibrational internal energy ($U_{L}$). 

\paragraph{}

We calculate the bulk dielectric constants using Density Functional Perturbation Theory (DFPT) as implemented in VASP. We first calculate the permittivity tensor of the bulk unit cell. We then obtain the in-plane and the out-of-plane dielectric constants from the permittivity tensor. We compute the static dielectric constant ($\epsilon_{0}$), which includes both the electronic and ionic responses. We also determine the optical dielectric constant ($\epsilon_{\infty}$) at high frequency, when only electrons respond to the external field while ions stay fixed in their lattice sites. To acquire the contribution of a monolayer itself, we subtract the vacuum contribution from the supercell dielectric constants and rescale the supercell dielectric constants using the following rescaling formula \cite{osanloo2021identification, laturia2018dielectric}:

\begin{equation}
 \epsilon_{2D,\perp} = [1+\frac{c}{t} (\frac{1}{\epsilon_{SC,\perp}}-1)]^{-1}\
\end{equation}

\begin{equation}
 \epsilon_{2D,\parallel} = [1+\frac{c}{t} ({\epsilon_{SC,\parallel}}-1)]\
\end{equation}

where $c$ and $t$ are the size of supercell and the monolayer thickness, respectively. 
\paragraph{}

To compute the thermionic and tunneling current densities through the metal-semiconductor, we use the direct tunneling equations \cite{gehring2003simulation, yeo2000direct}: 

\begin{equation} 
J_{\mathrm{tun}}=\frac{q^3\ {\mathcal{E}}^2}{8\pi h(\varphi-\varphi_{0})}{\mathrm{exp} \left(\frac{-4\sqrt{2m^*}\ {(\varphi }^{{3}/{2}}-\varphi_{0}^{{3}/{2}})}{3q\hbar\mathcal{E}}\right)\ } 
\end{equation}

\begin{equation} 
J_{\mathrm{therm}}=A^{**}\ T^2\ {\mathrm{exp} \left(\frac{-q\left(\varphi -\sqrt{\frac{q\mathcal{E}}{4\pi {\varepsilon }_i}}\right)}{kT}\right)\ } 
\end{equation} 

where $\mathcal{E}$, $\varphi $, ${\varepsilon }_i$, $A^{**}$, \textit{q}, $m^{\mathrm{*}}$, \textit{k}, and \textit{T} are the electric field in the insulator, barrier height, insulator permittivity, effective Richardson constant, electron charge, electron effective mass, Boltzmann constant, and temperature, respectively. In Eq.~3 and Eq.~4 $\varphi $ and $\varphi_{0} $ are the height of energy barrier so that $\varphi_{0} $ = $\varphi $ - $V_{\mathrm{g}}$, and  $V_{\mathrm{g}}$ =  $V_{\mathrm{DD}}$ - $V_{\mathrm{thr}}$. $V_{\mathrm{DD}}$ and $V_{\mathrm{thr}}$ are sequentially the supply voltage (0.7 V) and threshold voltage (0.345 V) as determined by IRDS in 2020 \cite{moore2020international}. To calculate the electric field inside the insulator we have:

\begin{equation}
\mathcal{E} = \frac{(V_{\mathrm{DD}} - V_{\mathrm{thr}})}{t}
\end{equation} 

where $t$ is thickness of monolayer shown in Fig~\ref{fig:structure} 
\paragraph{}

We calculate the electron effective mass in the out-of-plane direction as reported in Table~\ref{tab:leakage_current}. We derive the effective mass from the curvature of the bulk band structure by considering a 100 k-point path along the high symmetry path in out-of-plane direction using the PBE functional:

\begin{equation}
(m^{*})^{-1} = \frac {1}{\hbar^{2}} \frac{d^{2}E}{dk^{2}}
\end{equation} 
where $E(k)$ is the energy of the carrier and $k$ is the component of the wavevector in the out-of-plane direction, and $\hbar$ is the reduced Plank constant. The electron effective mass of \ch{Ca(OH)_{2}} and \ch{Mg(OH)_{2}} (out-of-plane tunneling mass) are computed to be 0.538 and 0.501, respectively. The effective mass of h-BN (0.47 $m_{e}$) is obtained from Ref \cite{palla2016bandgap}.

\section{Conclusion}
We investigated two novel 2D layered materials, \ch{Ca(OH)_{2}} and \ch{Mg(OH)_{2}}, for their potential applications as dielectrics in $n$-MOS devices. For each material, we calculated the exfoliation energy, band offset, phonon spectrum, thermodynamic properties, EOT, and leakage current. The exfoliation energies confirm that both materials are mechanically exfoliable and can be isolated in layers. The strictly positive phonon spectra clearly show the thermodynamic stability of the monolayers. We studied the thermodynamic properties and observed that the smaller free energy of \ch{Mg(OH)_{2}} makes its surface more favorable applications compared to  \ch{Ca(OH)_{2}}. In addition, the lower solubility of \ch{Mg(OH)_2} in water grants it a privilege over \ch{Ca(OH)_2} for industrial applications. We also used DFPT to calculate the in-plane and out-of-plane macroscopic dielectric constants. Although the in-plane static dielectric constant of monolayer \ch{Ca(OH)_{2}} is $~$15$\%$ higher than that of \ch{Mg(OH)_{2}}, in the out-of-plane direction a single-layer of \ch{Mg(OH)_{2}} has relatively a higher dielectric constant (6.40) than \ch{Ca(OH)_{2}} (6.33). We calculated the leakage current and the EOT for each material to evaluate its performance as a gate dielectric when combined with TMD channel materials. Our calculations show that bilayer \ch{Mg(OH)_{2}}/\ch{WS_{2}} heterostructure offers a lower leakage current compared to \ch{Ca(OH)_{2}}/\ch{HfS_{2}} heterostructure. While other layered dielectrics like LaOCl could provide an even better performance, \ch{Ca(OH)_2} and \ch{Mg(OH)_2}, are available both in nature and commercially synthesized on sizeable crystals.  Moreover, we validate that the predicted HSE bandgaps using Anderson's rule for the \ch{Ca(OH)_2}/\ch{HfS_2} heterostructure are in good agreement with the calculated HSE bandgap for \ch{Ca(OH)_2} and its associated TMD channel, \ch{HfS_2}. The band offset in \ch{Ca(OH)_{2}}/\ch{HfS_2} \ch{Mg(OH)_{2}}/\ch{WS_2} heterostructures using Anderson's rule are 1.20 eV and 1.99 eV, while we calculate a 0.97 eV, and 2.11 eV band offset for monolayer \ch{Ca(OH)_{2}} and \ch{Mg(OH)_{2}} when combined with \ch{HfS_2} and \ch{WS_2}, respectively. The calculated leakage current for 2L \ch{Ca(OH)_2} and \ch{Mg(OH)_2} are much lower than IRDS requirement and their 2L EOTs are calculated to be 0.56 nm and 0.60 nm, respectively. Our results show that a FET with a monolayer of \ch{Mg(OH)_{2}} and as a dielectric would outperform monolayer \ch{Ca(OH)_{2}} and h-BN. 

There are a few unanswered questions that need further clarification:

(i) Solubility is critical factor for different applications. Interestingly, at room temperature the solubility of \ch{Mg(OH)_{2}} ($\sim$ 9.80 $\times$ $10^{-4}$ g/ml) is very close to the solubility of amorphous \ch{SiO_{2}} ($\sim$ 1.2 $\times$ $10^{-4}$ g/ml) \cite{oswald1977bivalent, thilo1961einige}. In contrast to \ch{SiO_{2}}, due to the positive heat of solution in \ch{M(OH)_{2}} materials, their solubility decreases with increasing temperature \cite{oswald1977bivalent}. Although at room temperature the solubility of \ch{Ca(OH)_{2}} ($\sim$ 1.09 $\times$ $10^{-1}$ g/ml) is significantly higher than the solubility of amorphous \ch{SiO_{2}} ($\sim$ 1.2 $\times$ $10^{-4}$ g/ml), the solubility of both in water is the same at around 400 K \cite{oswald1977bivalent, arabi2015formation}. At higher temperatures \ch{Ca(OH)_{2}} has a lower solubility compared to amorphous \ch{SiO_{2}}. Hence, due to high level of reactivity of \ch{M(OH)_{2}} materials with water a proper encapsulation architecture is required to enhance the long-term stability of \ch{M(OH)_{2}} based devices. 

(ii) Although both \ch{Ca(OH)_{2}} and \ch{Mg(OH)_{2}} are stable against oxidation at room temperature, they are not stable against \ch{CO_{2}} \cite{oswald1977bivalent}. Consequently, these materials have been used for carbon capture and thermal heat storage. \ch{M(OH)_{2}} materials and other members of this family (\emph{i.e.,} \ch{Cd(OH)_{2}}, \ch{Ni(OH)_{2}}, \ch{Zn(OH)_{2}})  have the potential to contribute to the worldwide goal of decarbonization through carbon capture and storage, which would help to protect our planet from the catastrophic effects of climate change \cite{yuan2018cao,fagerlund2011experimental, piperopoulos2021tuning}.

(iii) We combined \ch{Ca(OH)_{2}} and \ch{Mg(OH)_{2}} with two TMD channels (\ch{HfS_{2}} and \ch{WS_{2}}) in this work; there are a handful of \ch{M(OH)_{2}}/TMD (\emph{i.e.,} \ch{Mg(OH)_2}/\ch{MoS_2}) combinations that need to be carefully investigated to find the most promising heterostructure candidates for \emph{p}-MOS and \emph{n}-MOS applications. The bandgap tuning of the heterostructures in the presence of an external field \cite{yagmurcukardes2019raman, yagmurcukardes2016mg}, the effect of strain and stress, as well as the defect formations in monolayers and bilayers, are other interesting topics for future studies.

We expect that our findings along with the recent crystalline synthesis of \ch{Ca(OH)_{2}}  and \ch{Mg(OH)_{2}} will lead to more research into the combination of novel 2D layered dielectrics with other prominent TMD channels for applications in 2D FETs.

\section*{Acknowledgement}
This research was sponsored in part by the Semiconductor Research Corporation (SRC) under the Logic and Memory Devices (LMD) Program of the Global Research Collaboration (GRC). This material is based upon work supported by the National Science Foundation under Grant No. 1802166. The project or effort depicted was or is sponsored by the U.S. Department of Defense, Defense Threat Reduction Agency. The authors acknowledge Christopher L. Hinkle for fruitful discussions.

\printbibliography
\end{document}